\documentstyle[12pt]{article}

\begin{document}

\vskip 10mm

\centerline{\Large \bf Comment on the Origin of the Supermassive Black Hole}

\vskip 20mm

\centerline{Kazuyasu Shigemoto \footnote{Department of Physics,
Tezukayama University, Nara 631-8501, Japan \\
\qquad E-mail address:shigemot@tezukayama-u.ac.jp} }

\centerline{\it Department of Physics}
\centerline{\it Tezukayama University, Nara 631-8501, Japan}

\vskip 10mm

\centerline{\bf Abstract}

We give the comment why the supermassive black hole 
exists at the center of almost all galaxies.  
We consider the origin of the supermassive black hole from the 
point of view of the density of the matter. 
If the density of the matter is fixed and such matter can 
come together by the gravitational attraction, such stellar object 
eventually becomes the black hole.
If the density of the matter is that of the atom, such matter 
naturally comes together to form the black hole by the gravitational 
attraction. This is expected to be the process to form the 
supermassive black hole. In this way, we can understand that 
there exists supermassive black hole at the center of almost 
all galaxies, because there is no delicate process to evolve into 
the black hole. 

\vskip 10mm

\section{Introduction}
\indent

Recently we have the observational evidence that there exists
the supermassive black hole at the center of almost 
all galaxies\cite{kormendy}\cite{melia}.
The mass of such supermassive black hole ranges from $10^6 M_{\odot}$
to $10^9 M_{\odot}$. It is well understand how the ordinary black hole 
is formed from the ordinary star \cite{brown}. But it is not established 
how the supermassive black hole at the center of the galaxies
is formed\cite{bekenstein}. Whether the stellar object becomes 
the black hole or not is related with the radius and the mass 
of that object. 
But we must more carefully study whether such stable state, especially 
the density of the matter, physically exists or not. For example, it 
is quite improbable that there exists the mini black hole with 
the radius of a few centimeters, because such stable state of 
that density is not physically known. 
Even if such stable state of the density exist,
we must study how to evolve into the black hole from the stellar
gas. 

In this paper, we study the black hole from the point of view whether 
such black hole is the stable state of the physically known density 
or not, and we study how the supermassive black hole is 
naturally formed from the stellar gas.

\section{Ordinary Black Hole}
\indent

We use Misner-Thorne-Wheeler notation \cite{Wheeler} and 
consider the Einstein equation of motion in the form  

\begin{eqnarray}
  R_{\mu \nu}-{1 \over 2} g_{\mu \nu} R 
  =\frac{8 \pi G T_{\mu \nu}}{c^4} , 
\label{e1} 
\end{eqnarray}
where $G$\ is the gravitational constant, $R$\ is the scalar 
curvature, $R_{\mu \nu}$ is the Ricci tensor and $T_{\mu \nu}$ 
is the energy-momentum tensor of the matter.
The metric of the Schwarzshild solution is given by 

\begin{eqnarray}
  ds^2=-(1-a/r)dt^2+\frac{1}{1-a/r}dr^2+r^2 \left(d\theta^2
       +\sin{\theta}^2 d\phi^2\right) ,
\label{e2} 
\end{eqnarray}
where $\displaystyle{a=\frac{2GM}{c^2}}$.

For the ordinary star, we have 
\begin{eqnarray}
  r_{\rm ordinary\ star}>>a=\frac{2GM}{c^2} ,
\label{e3} 
\end{eqnarray}
so that the ordinary star does not become 
the black hole. While, for the special star, which 
satisfies 
\begin{eqnarray}
  r_{\rm B.H.} \le a=\frac{2GM}{c^2} ,
\label{e4} 
\end{eqnarray}
it becomes the black hole. 
The values of the physical constants in the above 
are given by

\begin{eqnarray}
  &&G=6.67\times 10^{-11}\ m^3/kg \cdot s^2 , \nonumber\\
  &&c=3.0\times 10^{8}\ m/s . \nonumber\\
\label{e5} 
\end{eqnarray}

We use the following values of various physical quantities 

\begin{eqnarray}
  &&M_{\odot}=2.0\times 10^{30}\ kg ,\qquad
  r_{\odot}=7.0\times 10^{8}\ m , \nonumber\\
  &&M_{\rm earth}=6.0\times 10^{24}\ kg ,\qquad
  r_{\rm earth}=6.4\times 10^{6}\ m , \nonumber\\
  &&M_{\rm nuetron}=1.67\times 10^{-27}\ kg ,\qquad
  r_{\rm neutron}=10^{-15}\ m \nonumber\\
  &&M_{\rm hydrogen\ atom}=1.67\times 10^{-27}\ kg , \qquad
  r_{\rm hydrogen\ atom}=10^{-10}\ m . \label{e6} 
\end{eqnarray}

\subsection{Black hole radius for the sun}
\indent

We use the above physical values, and we evaluate the 
critical radius $r_{\odot\ {\rm B.H.}}$ for the sun.
If the radius of the sun becomes smaller than this 
critical radius, the sun becomes the black hole. 
This critical radius becomes   
 
\begin{eqnarray}
r_{\odot\ {\rm B.H.}}=\frac{2GM_{\odot}}{c^2}
=3\times 10^{3} \ m .
\label{e7}
\end{eqnarray}

\subsection{Black hole radius for the neutron star}
\indent

We put the radius of the neutron star as $L$ times 
the neutron radius, that is, 
$r_{\rm neutron\ star}=L r_{\rm neutron}$,
then the mass of the neutron star is given by 
$L^3 M_{\rm neutron}$. The critical condition that 
the neutron star becomes the black hole is given by

\begin{eqnarray}
L r_{\rm neutron}=\frac{2GL^3 M_{\rm neutron}}{c^2} ,
\label{e8}
\end{eqnarray}
which gives 

\begin{eqnarray}
L=\sqrt{\frac{r_{\rm neutron}c^2}{2GM_{\rm neutron}}}=
2\times 10^{19} .
\label{e9}
\end{eqnarray}
Then we have the typical mass of the neutron black hole, 
ordinary black hole, as 

\begin{eqnarray}
M_{\rm neutron\ B.H.}=L^3\times M_{\rm neutron}
=1.34\times 10^{31}\ kg= 6.7\times\ M_{\odot} , 
\label{e10}
\end{eqnarray}
which agrees with the observation.
The typical radius of the neutron black hole is 
\begin{eqnarray} 
r_{\rm neutron\ B.H.}=L \times r_{\rm neutron}
=2\times 10^{4}\ m .
\label{e11}
\end{eqnarray}

\section{Supermassive Black Hole}
\subsection{Black hole condition from the fixed density of the matter}
\indent

We denote the density of the matter as $\rho$, then we have 
$ \displaystyle{M=\frac{4 \pi r^3 \rho}{3}}$.
The critical radius that the stellar object becomes 
the black hole is expressed as
$\displaystyle{a=\frac{2GM}{c^2}=\frac{8 \pi G \rho}{3 c^2} r^3}$.
Then if we take the radius $r$ to be large enough, we eventually have 
$\displaystyle{r < a=\frac{8 \pi G \rho}{3 c^2}r^3}$. 
Therefore, if the matter is in the stable state with the 
physical density and such matter can come together by 
the gravitational attraction, it eventually evolve to the black hole.

\subsection{Black hole radius for the stellar object with hydrogen atom}
\indent

We put the radius of the stellar object with hydrogen atom
as $L$ times that of the hydrogen atom,
that is, $r_{\rm hydrogen\ star}=L r_{\rm hydrogen}$,
then the mass of the stellar object with hydrogen atom
is given by 
$L^3 M_{\rm hydrogen}$. The condition that the 
the stellar object with hydrogen atom becomes the 
black hole is given by

\begin{eqnarray}
L r_{\rm hydrogen\ atom}=\frac{2GL^3 M_{\rm hydrogen\ atom}}{c^2} ,
\label{e12}
\end{eqnarray}
which gives 

\begin{eqnarray}
L=\sqrt{\frac{r_{\rm hydrogen\ atom}c^2}{2GM_{\rm hydrogen\ atom}}}=
6.3 \times 10^{21} .
\label{e13}
\end{eqnarray}
Then we have the typical mass of the black hole with 
the hydrogen atom
as 
\begin{eqnarray}
M_{\rm hydrogen\ B.H.}=L^3\times M_{\rm hydrogen\ atom}
=4.2\times 10^{38}\ kg =2\times 10^{8}\ M_{\odot} 
\label{e14}
\end{eqnarray}
This mass is in the range of the observed mass of the 
supermassive black hole $10^6 M_{\odot} \sim 
10^9 M_{\odot}$. Then we expect that the supermassive black 
hole is the black hole composed of the hydrogen atom.
The typical radius of the black hole with hydrogen atom is  
given by 

\begin{eqnarray} 
r_{\rm hydrogen\ B.H.}=L \times r_{\rm hydrogen\ atom}
=6.3\times 10^{11}\ m=4.3 \times {\rm cosmological\ unit}.
\label{e15}
\end{eqnarray}

\section{Summary and Discussion}
\indent

We study why the supermassive black hole exists at the center 
of almost all galaxies. As the supermassive black hole exists 
at almost all galaxies, it must exist the quite natural process 
of the stellar gas evolving into the supermassive black hole. 
We study in this paper from the point of view whether the density 
of the black hole is the physically stable state or not.
There are two typical stable density, that is, the nucleus density
and the atomic density. It is the quite delicate process to 
to reach the nucleus density state from the stellar gas.
The stellar gas forms the heavy star and such heavy star collapse,
then we have the ordinary black hole if some delicate condition 
is satisfied.
While, if the density of the matter is that of the atom, the stellar gas 
can come together by the gravitational attraction without any delicate 
condition. Then such object eventually becomes large enough and
we naturally have the supermassive black hole. In this approach, 
we estimate the typical mass of the supermassive black hole as 
$10^8 M_{\odot}$, which is in the observational range 
of the supermassive black hole $10^6 M_{\odot} \sim 10^9 M_{\odot}$ . 


\end{document}